\documentclass[preprint,12pt]{elsarticle}

\usepackage{subfigure}
\usepackage{pstricks}
\usepackage[percent]{overpic}
\usepackage{color}
\usepackage{rotating}
\usepackage{tabularx}

\usepackage{amssymb}
\usepackage{amsmath,amsfonts,amsthm}

\journal{Transportation Research Part B: Methodological}

\begin{document}

\begin{frontmatter}



\title{Dynamic Stride Length Adaptation According to Utility And Personal Space}


\author{Isabella von Sivers\corref{cor1}}
\ead{isabella.von\_sivers@hm.edu}
\author{Gerta K\"oster\corref{}}
\ead{gerta.koester@hm.edu}
\cortext[cor1]{Corresponding author, Tel.: +49-89-1265-3762}

\address{Munich University of Applied Sciences, Lothstra{\ss}e 64, 80335 Munich, Germany}

NOTICE: this is the author’s version of a work that was accepted for publication in Transportation Research Part B: Methodological. Changes resulting from the publishing process, such as peer review, editing, corrections, structural formatting, and other quality control mechanisms may not be reflected in this document. Changes may have been made to this work since it was submitted for publication. A definitive version was subsequently published in Transportation Research Part B: Methodological, 2015, Vol. 74, pages 104 -- 117, DOI: 10.1016/j.trb.2015.01.009

\begin{abstract}

Pedestrians adjust both speed and stride length when they navigate difficult situations such as tight corners or dense crowds. They try to avoid collisions and to preserve their personal space.
State-of-the-art pedestrian motion models automatically reduce speed in dense crowds simply because there is no space where the pedestrians could go. The stride length and its correct adaptation, however, are rarely considered. This leads to artefacts that impact macroscopic observation parameters such as densities in front of bottlenecks and, through this, flow.  Hence modelling stride adaptation is important to increase the predictive power of pedestrian models.
To achieve this we reformulate the problem  as an optimisation problem on a disk around the pedestrian. Each pedestrian seeks the position that is most attractive in a sense of balanced goals between the search for targets, the need for individual space and the need to keep a distance from obstacles. The need for space is modelled according to findings from psychology defining zones around a person that, when invaded, cause unease.
The result is a fully automatic adjustment that allows calibration through meaningful social parameters and that gives visually natural results with an excellent fit to measured experimental data.  
\end{abstract}

\begin{keyword}
Pedestrian Movement, Crowd Dynamics, Optimal Steps Model, Stride Length Adaptation, Personal Space,
Optimisation, Nelder Mead Simplex
\end{keyword}

\end{frontmatter}



\section{Introduction}  

Simulation of pedestrian movement becomes increasingly important to ensure safety for everybody wherever a crowd comes together. Large and dense crowds shape daily urban traffic, not only in areas assigned to walking but also at transfer locations such as train platforms or on shared spaces \cite{kretz-2013}. Thus, pedestrian movement is an integral part of the overall urban transportation problem and adequate modelling of human locomotion in a crowd has become essential to its solution.

Many new models for microscopic pedestrian transport have been presented over the last years and existing models are being constantly refined \cite{gwynne-1999,robin-2009,zheng-2009,smith-2009,asano-2010,seitz-2012}. Among them cellular automata \cite{burstedde-2001,blue-2001,kirchner-2003,henein-2007,ezaki-2012,kneidl-2013} and social force models \cite{helbing-1995,helbing-2000,chraibi-2010,koster-2013} are prominent and, perhaps, best investigated. But some typical aspects of human movement are still missing in known models, in particular the immediate adaptation of the stride length to the navigational situation.
Yet stride length adaptation is closely connected to at least the latter two of the four most prominent `process variables' of pedestrian traffic listed in \cite{daamen-2003}: the free-flow velocity, the walking direction, the crowd density, and the effect of bottlenecks.

When pedestrians navigate a difficult corner or walk within a crowd they reduce their speed. And more than that, they make smaller steps. They do this with foresight, that is, they adapt to the situation before or at the moment they encounter it. Avoiding collisions is clearly one reason for this. Another one is that pedestrians try to preserve their personal space \cite{katz-1937,sommer-1959,hall-1966} when approaching others who they do not identify with \cite{novelli-2013} adding a social aspect to their behaviour. The psychological model of personal space introduced by Hall in 1966 \cite{hall-1966} is widely accepted and built upon, e.g. \cite{beaulieu-2004,uzzell-2006,evans-2007,kennedy-2009}.

Following \cite{seyfried-2010}, smaller strides in dense situations are, at least, one reason for speed reduction. However, most current state-of-the-art pedestrian motion models are only capable of speed adaptation, typically in a reactive way, but do not adjust the stride length. In fact, standard continuous models, based on differential equations, do not even model human steps but rather a smooth sliding motion of particles. Step sizes refer to numeric advances in time without any bio-mechanical meaning.  
Cellular automaton models only allow `hops' from one cell to the next. Stride adaptation is impossible and the choice of direction is very limited. 

Clearly, models of personal space that demand a fine spatial resolution do not make much sense for cellular automata. However, the concept is also lacking in most of the spatially continuous models of pedestrian traffic. Sterlin et al. \cite{sterlin-2010} briefly show how it fits into an agent-model but do not give any information on their choice of locomotion model nor simulation results. In robotics, on the other hand, the concept of `personal space' is widely used. The distance robots keep from humans is mostly based on personal spaces according to \cite{hall-1966}. They adopt the psychological model to let the robots mimic natural human behaviour -- to keep distances from others. The vast variety of articles on `personal space' in robotics cannot be covered here. Important advances in research can be found, e.g., in \cite{tasaki-2004,pacchierotti-2005,walters-2009,mead-2013,vasquez-2013}.

The goal of this paper is to develop a model that fully models step-wise movement with immediate stride length adaptation based on the need to keep a distance from others -- as observed by psychologists \cite{hall-1966} -- and from obstacles (e.g. \cite{seitz-unpublished}). We argue, that the model matches human walking much more closely and its calibration parameters have a social meaning.  
The model leads to visually more natural results and a good fit with measured data from well-known experiments (see Section \ref{validation}).

For this, we further develop the Optimal Steps Model (OSM), one of the newer approaches to modelling pedestrian dynamics introduced in \cite{seitz-2012}. We choose the OSM because it already models stepwise movement in arbitrary directions.
The former state of the OSM described in \cite{seitz-2012} allows step-wise movement and even stride adaptation, but only in a reactive way, tying the stride length to the actual speed of movement. This makes sense in a free-flow situation where there is a linear dependency on the walking speed as shown in \cite{seitz-2012} and previously investigated e.g. in \cite{kirtley-1985,grieve-1966}. In this paper, however, we are not looking into free-flow situations so that the relationship from \cite{seitz-2012} need no longer hold. 
In addition, speed can only be measured using positions from the past which introduces a delay in reaction if the stride-length is modelled as a function of speed. Then pedestrians emerging from a dense crowd accelerate but begin with small steps instead of long strides. On the other extreme end, large steps when walking into a crowd or approaching a tight gap make navigational fine tuning difficult, even when the speed is low. Agents may get stuck where real pedestrians still move \cite{sivers-2013b} and the flow may be reduced to zero where it should not be. Hence we need a modelling concept that allows to immediately and simultaneously adapt speed and stride length.

Our ansatz is simple: We search the next position of the pedestrian on a disc instead of the circle around the pedestrian. The radius of the disc is the stride length corresponding to the pedestrian's free-flow speed. Within the disc, utility is given by a balance between closeness to the target, and distance to obstacles and maintained distance to others. We obtain a new two-dimensional optimisation problem which is described in a mathematically more rigorous way in \cite{sivers-2013}. In this paper we additionally exploit the new fine resolution of steps and spacial use to implement an empirically substantiated concept of personal space \cite{hall-1966} that is well accepted among psychologists. Utility is coded in a floor field, again building on top of a very successful modelling technique which was introduced to pedestrian dynamics, to our knowledge, by \cite{burstedde-2001} and adopted and altered by many, e.g. \cite{klupfel-2003,nishinari-2004b,kirik-2009,hartmann-2010,koster-2011,bandini-2014,hartmann-2014,koster-2014b}.

The paper is structured as follows: Section \ref{model} introduces the enhanced OSM and its tactical navigation using functions to express utility, in particular, utility of preserved personal space. In 
Section \ref{numerical_solution}, we formulate stepping ahead mathematically as a two-dimensional optimisation problem and demonstrate how to solve it numerically. This problem has a potential for high computational cost. Hence an efficient numerical treatment is necessary. We will also tackle this in section \ref{numerical_solution}.
In Section \ref{validation}, we show simulation results. We demonstrate that we are able to calibrate to any given density-speed relationship and how the new stride length adaptation helps to capture the density distribution in front of bottlenecks. In section \ref{conclusion} we discuss the results and give an outlook.

\section{Navigation in the enhanced Optimal Steps Model \label{model}}

\subsection{The concept of optimizing utility within the pedestrian's reach}

The OSM is inspired by the rule based approach of cellular automata models, in particular the rules of the model described in \cite{koster-2011b}. However, movement is not restricted to a grid. In the Optimal Steps Model in its original form \cite{seitz-2012} the next position of a pedestrian is chosen on a circle with a fixed radius around the pedestrians midpoint. The radius of the circle represents the stride length. 

Pedestrians seek targets and avoid other pedestrians and obstacles. They navigate along a floor field constructed by superposing scalar fields. The three scalar functions express the orientation towards a target, the need to avoid too close contact with fellow pedestrians and the necessity to skirt obstacles. 

The traditional way to look at these functions is to think of them as potentials that, through the potentials' gradients, represent forces between the target, the pedestrians and the obstacles \cite{helbing-1995,molnar-1996}. This inspiration from physics has produced a model generation including social force models. However, in reality, there are no such forces. 
Thus, in the OSM we prefer to follow the idea of a utility function. The idea of fully rational humans optimizing utility has been criticized in favour of simple rule-type decisions \cite{gigerenzer-2008b, moussaid-2011}. However, we still find it plausible that persons seek advantageous positions. Instead of using the gradient to determine the velocity and thus, indirectly, the direction of movement as in social force models the OSM finds the next position by maximizing the utility or, equivalently, by minimizing the potential.

In this paper, we replace the former optimisation on a circle around the pedestrian by optimisation on the full disc \cite{sivers-2013}. 
Now, every position within the step circle of a person becomes possible; position seeking can be fine tuned. The radius $r_i$ of the disc, that is the maximum stride length of pedestrian $i$, is chosen in accordance with the free-flow velocity of pedestrian $i$. There is a mostly linear inter-dependency of stride length and free-flow speed that has been empirically shown, e.g, in \cite{grieve-1966,seitz-2012}. In our simulations, the free-flow speeds are chosen to be normally distributed. This is a wide-spread assumption that is backed by field observations, e.g., in \cite{davidich-2013}, and literature, e.g. \cite{weidmann-1993}.

With the ability to finely adjust positions in space, it makes sense to introduce a more complex concept of personal space than in our former models that takes into account findings from psychology: Hall's model of social space as introduced in \cite{hall-1966} and further examined in \cite{beaulieu-2004,uzzell-2006,evans-2007,kennedy-2009}. The realisation in the OSM will be described in the next section.

\subsection{Personal space in the Optimal Steps Model} 

The concept of personal space has a long history. In the 1930th, the idea of a personal space or aura -- carried around a human -- was first mentioned \cite{stern-1930,katz-1937}. In the following years and decades, researchers conducted studies and experiments of personal space and distance (e.g. \cite{sommer-1959,horowitz-1964,hall-1966,sussman-1978,severy-1979,strube-1984,beaulieu-2004,uzzell-2006,evans-2007,kennedy-2009,novelli-2010}). The personal distances, that is, the distances persons keep from others to feel comfortable measured in these experiments mainly vary between 45 and 120 cm depending on the experimental conditions, measurement method and the focus of the experiment. The distances also change with culture, age, gender, nationality, and identification but -- to the authors' best knowledge -- they remain in the range proclaimed by Hall \cite{hall-1966}. As a result, we follow and implement Hall's model.

\begin{figure}
\centering
	\subfigure[Spaces around a person according to \cite{hall-1966}. The (blue) filled circle in the middle represents the person.]{
	\label{fig:hall1}\includegraphics[height = 5.2cm]{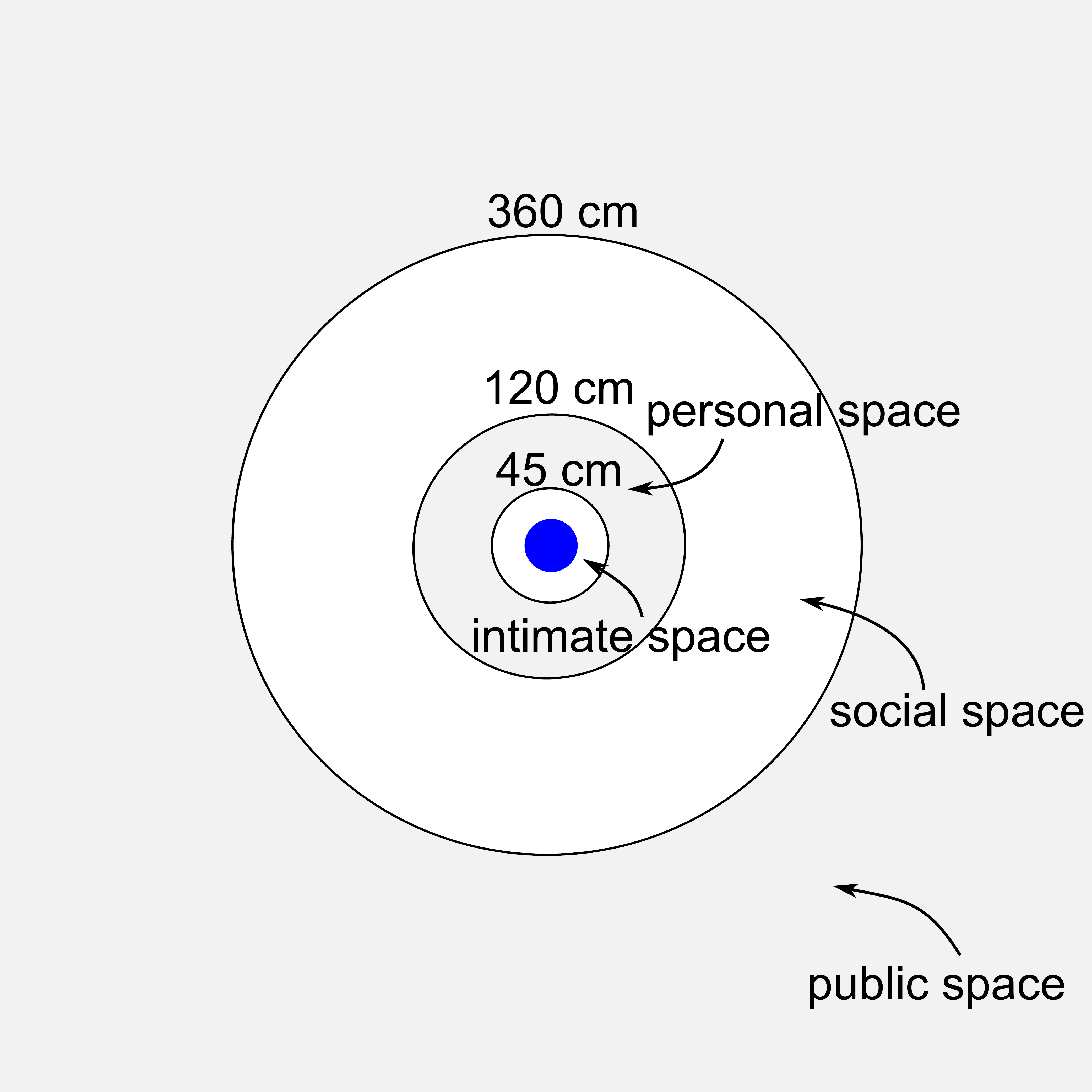}}
	\hspace{0.5cm}
	\subfigure[Close and far areas within the spaces according to \cite{hall-1966}. The closer areas in each space are filled in light grey.]{
	\label{fig:hall2}\includegraphics[height = 5.2cm]{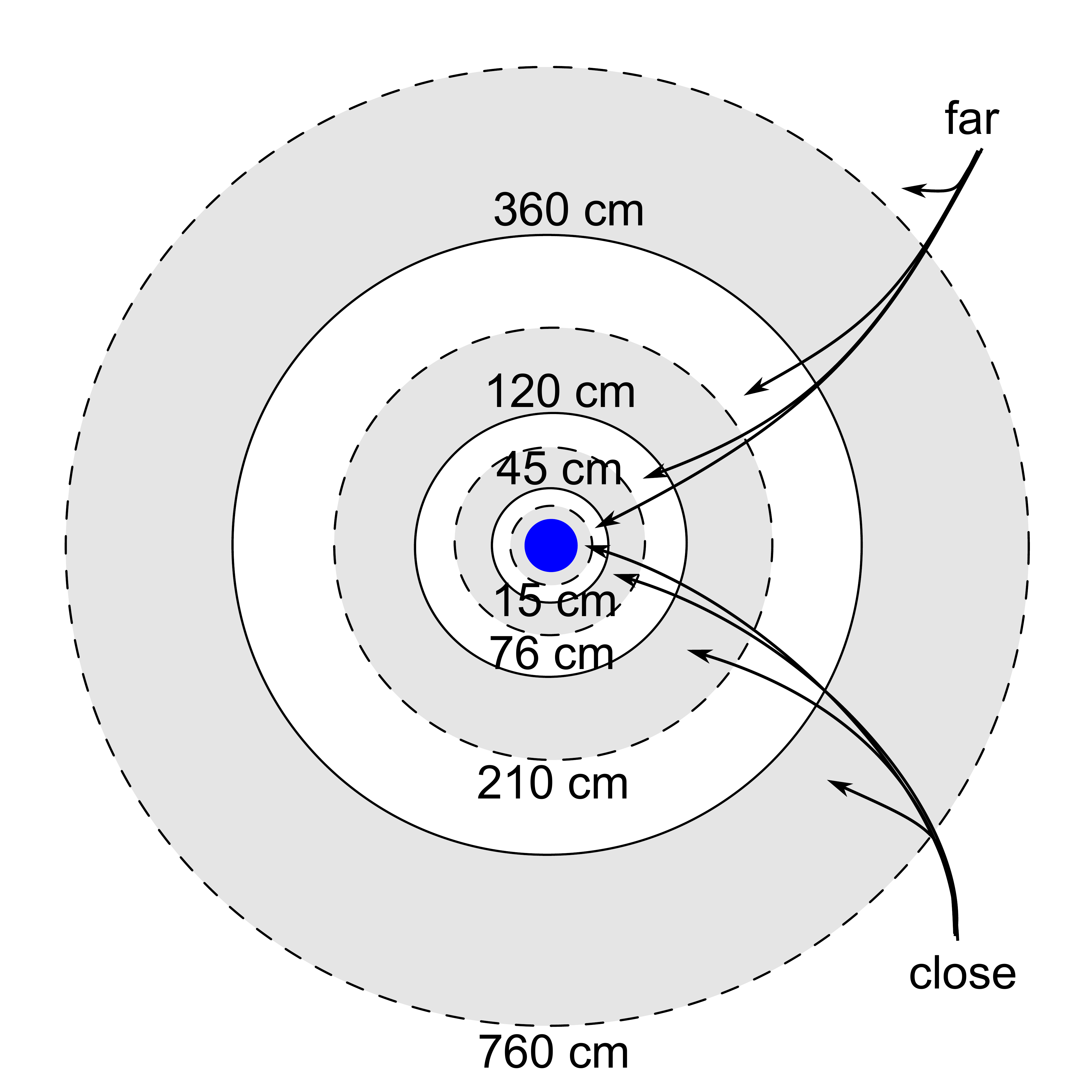}}
	\caption{The distances persons carry around them according to Hall's model \cite{hall-1966}.}
	\label{fig:hall}
\end{figure}

Hall's model of personal space was published in 1966 \cite{hall-1966}. Hall claimed that people have a special need for space and backed this by observations and interviews \cite{hall-1966}. He developed a model of four circular spaces that are carried around by every person: The intimate, personal, social, and public space (Fig.~\ref{fig:hall1}). Each of these spaces consists of a close and a far phase (Fig.~\ref{fig:hall}). The intimate space ($<$ 45 cm) defines the area where the sensory inputs (body heat, smell, sound, ...) of another person entering this area are intense and body contact is nearly unavoidable. Its close phase is reserved for sexual partners and children that are cared for and protected. The far phase can be entered by close friends to communicate but strangers are usually not accepted. In dense situations when strangers enter this zone, people tend to show defensive behaviour, like moving away or positioning the arms in a way to protect the body from contact as much as possible. The personal space (45 cm -- 120 cm) contains the distance where family and friends are accepted and the individual distance kept between a person and strangers. In its close phase people have `elbow room'. Its far phase is the area where even strangers are accepted: `keeping people at arm's length' \cite{hall-1966}. The beginning of the social space (120 cm -- 360 cm) marks the `limit of domination' \cite{hall-1966}. The social space is followed by the public space ($>$ 360 cm).

For pedestrian dynamics, the social and public space have little importance. According to Hall \cite{hall-1966}, there is nearly no direct effect on another person beyond 120 cm. Thus, we do not implement the social and public spaces in our model. We follow Hall's suggestions of circular intimate and circular personal spaces. For simplicity, pedestrians are also considered to be circles -- with radius $r_p$ = 20 cm. Other shapes could be modelled easily. 

\subsection{Utility functions and parameter choices}
 
To get the `target orientation' we look at a wave front propagating from the target as suggested in \cite{hughes-2000,treuille-2006,kretz-2009,kretz-2010,hartmann-2010}. Let $\Omega \subset \mathbb{R}^2$ be the area of observation of the scenario with boundary $\partial\Omega$. $\Gamma \subset \partial\Omega$ denotes the boundary of the target. Then the eikonal equation (\ref{eqn:eikonal}) describes this propagation through space starting from $\Gamma$. Its
solution $\Phi$ gives the arrival time $\Phi:\Omega \rightarrow \mathbb{R}$ of the wave front at a certain point in space.  The travelling speed of the front is given by $F: \Omega \to \mathbb{R}_{+}$. Typically, $F$ is chosen to be $1$ outside obstacles making the arrival time independent of surface conditions. More difficult terrain can be mapped by varying $F$ \cite{hartmann-2010}, as can be crowd avoidance \cite{kretz-2011,hartmann-2014b,koster-2014b} or, in fact, simple queuing. See \cite{koster-2014b} for queuing and a typical queue in Fig.~\ref{fig:bottleneck_density}. 
We assume that each pedestrian tries to choose the path with the shortest travel time to the target which amounts to to finding the location with the minimal arrival time $\Phi$ or, equivalently in terms of utility, maximizing $-\Phi$. 

The eikonal equation is given by
\begin{equation}
 \label{eqn:eikonal}
	F(x) \lVert \nabla \Phi(x)\rVert = 1 \hspace{1cm} \text{ for } x \in \Omega 
\end{equation}
with boundary condition
\begin{equation}
 \label{eqn:eikonalRB} 
	\Phi(x) = 0 \hspace{2cm} \text{ for } x \in \Gamma .
\end{equation}

Its numerical solution $\tilde{\Phi}(x)$ is efficiently computed by Sethian's fast marching algorithm on a two-dimensional grid \cite{sethian-1996,sethian-1999}. We interpolate $\tilde{\Phi}(x)$ bilinearily between grid points \cite{hartmann-2010,seitz-2012} to define the `target orientation' $P_t(x)$ for every point $x \in \Omega$. 

\begin{figure}
\centering
	\subfigure[Personal spaces inspired by \cite{hall-1966} around a person centered at the origin.]{
	\label{fig:pedPot1}\includegraphics[height = 5.2cm]{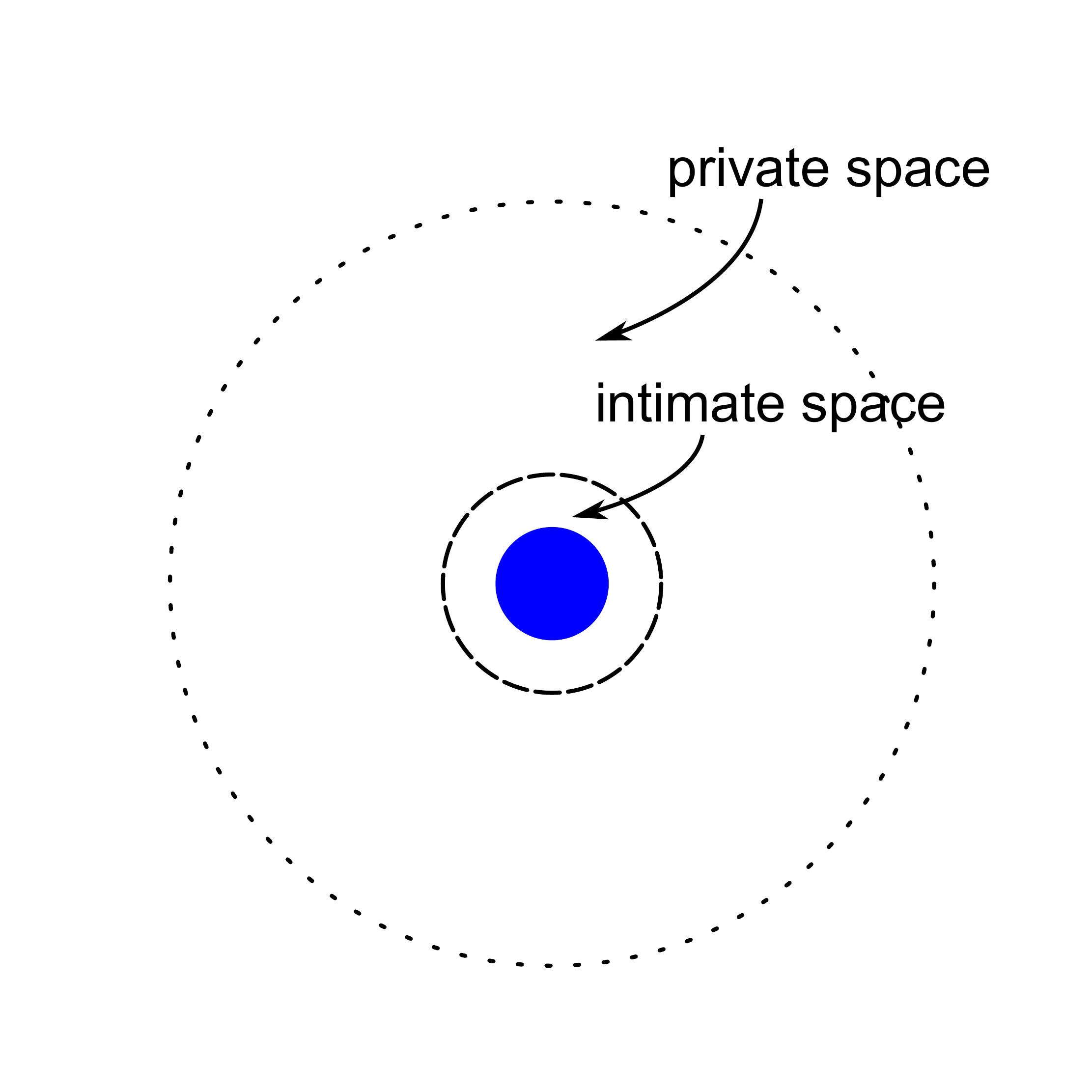}}
	\hspace{0.5cm}
	\subfigure[Illustration of a possible `pedestrian avoidance' function; the pedestrian is centered at the origin.]{
	\label{fig:pedPot2}\includegraphics[height = 5.2cm]{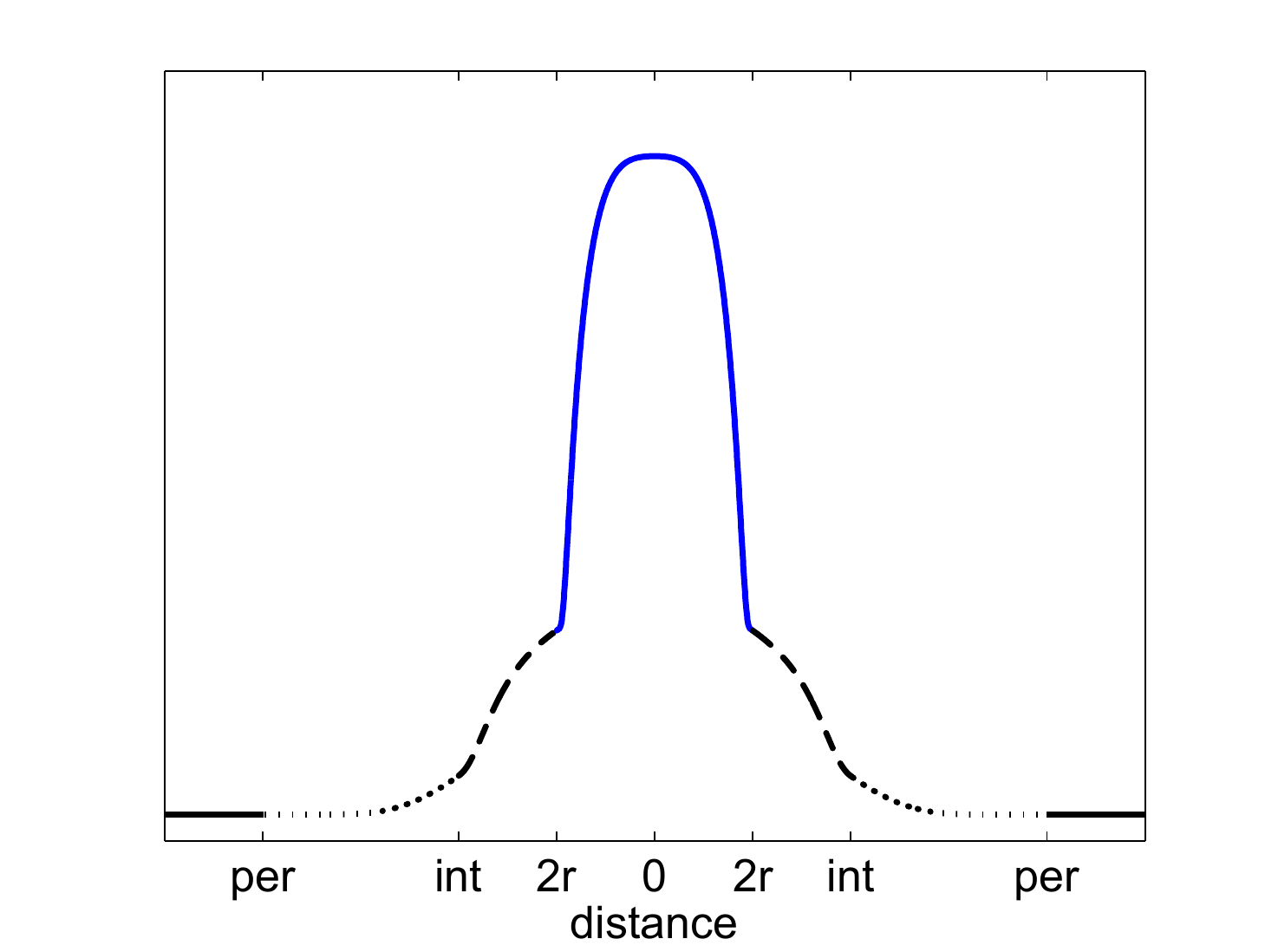}}
	\caption{The `pedestrian avoidance' in the Optimal Steps Model.}
	\label{fig:pedPot}
\end{figure}

In addition to the static `target orientation', the floor field contains dynamic `pedestrian avoidance' and `obstacle avoidance'.  
The `pedestrian avoidance' determines how strong the effect of a pedestrian on another one is. We define $p^j_1, p^j_2$ and $p^j_3$ to describe the need for personal space, intimate space and the torso representation (see Fig.~\ref{fig:pedPot1} of a pedestrian: 

\begin{align} 
 \label{eqn:perspotdefine} 
	\begin{array}{rl}	
			p^j_1(x) := & \mu_p \cdot \text{ exp}\left(\frac{4}{(d_j(x)/(\delta_{per}+r_p))^2-1}\right),\\
			p^j_2(x) := & p_1 + \frac{\mu_p}{a_p} \cdot \text{ exp}\left(\frac{4}{(d_j(x)/(\delta_{int}+r_p))^{2*b_p}-1}\right),\\
			p^j_3(x) := & p_2 + 1000 \cdot \text{ exp}\left(\frac{1}{(d_j(x)/r_p)^2-1}\right) 		
		\end{array}. &&&&
\end{align}

Note that we use smooth functions on compact support motivated by \cite{dietrich-2014} for the dynamic avoidance functions instead of functions of type $e^{-x^2}$. They truly vanish outside the personal space and there is no numerical cut-off error. 

The `pedestrian avoidance' of pedestrian $j$ is composed of the three functions from Eq. \ref{eqn:perspotdefine} (see Fig.~\ref{fig:pedPot1})

\begin{align} 
 \label{eqn:perspot} 
	P^j_p(x):=\left\{\begin{array}{ll}	
				p^j_3(x) \hspace{1cm} & d_j(x) < 2r_p,\\ 
				p^j_2(x) & 2r_p \leq d_j(x) < \delta_{int}+r_p,\\ 
				p^j_1(x) & \delta_{int}+r_p \leq d_j(x) < \delta_{per}+r_p,\\ 
				0 & \mbox{else}. 
		\end{array}\right. &&&&
\end{align}

While $P^j_p$ expresses the need for personal space of pedestrian $j$, in our model, it is the utility of the other pedestrians that is influenced by it. The outcome is the same. Each pedestrian tries to stay out of the personal space of the others. 

The value of the `pedestrian avoidance' depends on the Euclidean distance $d_j(x)$ between the centre of pedestrian $j$ and the considered position $x$ in the area. The function is structured in four areas 
(see Fig.~\ref{fig:pedPot1}): The first one is the overlapping area. 
The ring with outer radius $\delta_{int}$ around the overlapping area represents the intimate space.
It is followed by the personal space, a ring with outer radius $\delta_{per}$.
The slope of  the functions $p^j$ models the gradual transition from close phase to far phase in each space.
\\
The radii must be chosen according to experiments and/or existing literature. Our own choice  for all simulations in this paper is motivated by Hall's measurements \cite{hall-1966}: 
\begin{equation}
 \label{eqn:parametersPedPot} 
	\delta_{int} = 0.45m \text{ and } \delta_{per} = 1.20m.
\end{equation}
The only remaining parameters that still need to be calibrated are $\mu_p$ for the strength of the function, $a_p$ for the moderation of the strength between the intimate and the personal space and (the integer) $b_p$ for the slope of the function at the boundary of the intimate space. According to psychological findings, these parameters should change with culture, age, gender, and nationality \cite{hall-1966,sussman-1978,severy-1979,uzzell-2006}. Density-speed relationships are known to express the same variations in the crowd population \cite{hankin-1958,older-1968,weidmann-1993,lam-1995,chattaraj-2009,davidich-2013}. 
Thus, it makes sense to calibrate the social parameters $\mu_p, \, a_p, \, b_p$ by fitting the density-speed-relationship of the simulation output to a curve given by an experiment or field observations. We do this in Section \ref{subsec:density-speed}.

The `obstacle avoidance' $P^k_o$ 
\begin{align} 
 \label{eqn:obspot} 
	P^k_o(x):=\left\{\begin{array}{ll}	
				o^k_2(x) \hspace{1cm} & d_k(x) < r_p,\\ 
				o^k_1(x) & r_p \leq d_k(x) < \delta,\\ 
				0 &  \mbox{else}, 
		\end{array}\right. &&&&
\end{align} 
with 
\begin{align} 
 \label{eqn:obstpotdefine} 
	\begin{array}{rl}	
			o^k_1(x) := & \mu_o \cdot \text{ exp}\left(\frac{2}{(d_k(x)/(\delta_o))^2-1}\right),\\
			o^k_2(x) := & o_1 + 100000 \cdot \text{ exp}\left(\frac{1}{(d_k(x)/r_p)^2-1}\right)
		\end{array}, &&&&
\end{align}
is handled similarly. The distance $d_k(x)$ is the shortest distance between the boundary of the obstacle and the pedestrian's position $x$. According to experiments in \cite{seitz-unpublished}, the value of the `obstacle avoidance' depends on a `preferred distance' $\delta_o$ to the obstacle. Here, the only variable parameter is $\mu_o$. This parameter is calibrated in a way, that pedestrians still pass small corridors but keep the preferred distance when there is enough space. Thus, we get the following parameters:  
\begin{equation}
 \label{eqn:parametersObstPot} 
	\delta_o = 0.8m \text{ and } \mu_o = 6.0.
\end{equation}
 
For the reader's convenience, the parameters for the `pedestrian avoidance' and `obstacle avoidance' are described and listed in Table \ref{tbl:parametersEXP} and Table \ref{tbl:parametersCAL}.

For each pedestrian $i$ the superposition 
\begin{equation}
 \label{eqn:totalpot} 
	P_i(x) = P_t(x)+\sum^n_{j=1,j\neq i}{P^{j}_{p}(x)}+\max_{k=1...m}{P^{k}_{o}(x)}, 
\end{equation}
constitutes the floor field $P_i(x)$ for any point $x \in \Omega$. 

We interpret the floor field, or rather its negative, as a utility function or an objective function for 
pedestrian $i$. The closer to the target and the farther away from obstacles and other pedestrians the more attractive a position becomes. Function $-P$ measures the degree of the attraction, the utility, or $P$ the degree of discomfort.  
We now require that, with each step, each pedestrian selects the position with the lowest value of the floor field within his or her reach. That is, we search the next position on a disc.

\section{Mathematical formulation and numerical solution \label{numerical_solution}}

{These assumptions lead} to a two-dimensional optimisation problem with one inequality constraint:
\begin{align} 
 \label{eqn:optkreisscheibe2D} 
\begin{alignedat}{2}
	& &\min_{x \in \Omega}{P_i(x)}\hspace{2cm}\\
	&\text{s.t.} \hspace{2cm}& d_{i}(x)-r_i \leq 0 \hspace{2cm}.
\end{alignedat}
\end{align}
Here, $d_i(x)$ is the distance between the position of the considered pedestrian $i$ and the position $x$.

To find the minimum on the disc, we currently use two methods in our simulator: a straight forward search on grid points and the Nelder-Mead simplex algorithm \cite{nelder-1965}. We also experimented with evolutionary algorithms \cite{sivers-2013,kharchenko-2014}. 
The grid for the straight forward search method is constructed on concentric circles around the centre of a pedestrian. With a given tolerance $tol$ and the radius of the stride length $r$, the number of circles $c_c$ can be determined, according to \cite{sivers-2013}, by 
\begin{equation}
c_c := \left \lceil \frac{r}{\sqrt{2} \cdot tol} \right \rceil.
\end{equation}
The number of grid points $gp_i$ on circle $i$ with $1 \leq i \leq c_c$ is calculated through
\begin{equation}
	gp_i := \left \lceil \frac{i}{c_c} \cdot \frac{\sqrt{2}\pi r}{tol} \right \rceil.
\end{equation}
This assures that all grid points are evenly spaced on the whole disc. This method is mostly used when flexibility and stability is the goal and not accuracy of the solution.
The second method, the algorithm by Nelder and Mead, is a local direct search method for unconstrained optimisation that can deal with all kinds of functions, e.g., where no gradient information is available. It is usually applied to minimization problems. For functions on $\mathbb{R}^n$, the algorithm starts with an initial simplex $S_0$ constructed by $n+1$ vertices $x_0,...,x_n$. The Nelder-Mead method modifies and improves the simplex in every loop using four operations: reflection, expansion, contraction, and shrinkage. Mathematical details for our problem can be found in \cite{sivers-2013,sivers-2013b}. Convergence of the Nelder-Mead algorithm is sensitive to the choice of the initial simplex and cannot be guaranteed in general but, in practice, it provides good results in low dimensions regardless of the objective function \cite{lagarias-1998}. This is confirmed by our results \cite{sivers-2013}.

\begin{figure}
	\centering
		\includegraphics[width=\textwidth]{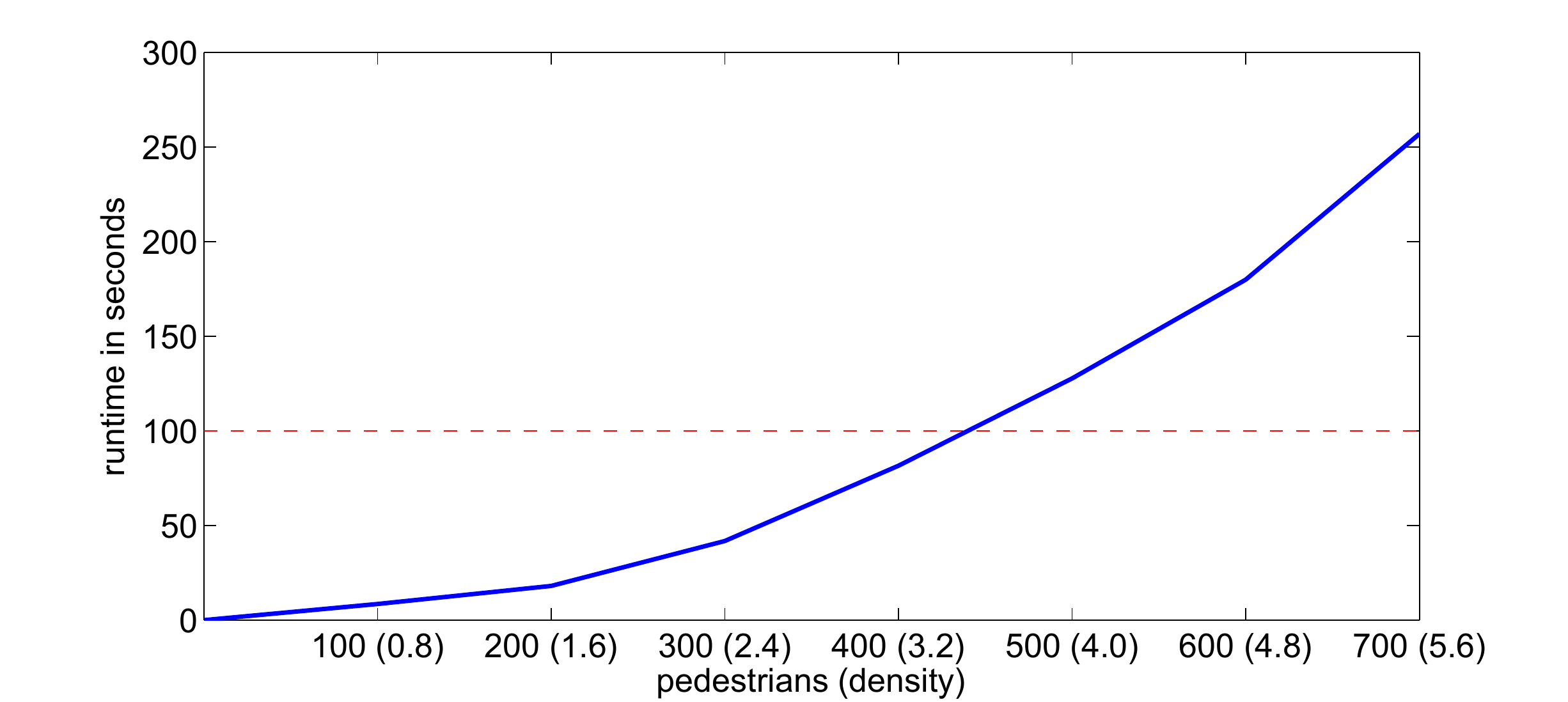}
		\caption{Simulation run times in a long corridor. Simulation of 100 real seconds with increasing numbers of pedestrians. The horizontal (red) dashed line shows the real time border that is crossed at about 430 pedestrians.}
	\label{fig:timemeasurement}
\end{figure}

When we compare the Nelder-Mead method to a straight forward search on a grid and also to evolutionary algorithms Nelder-Mead proves to be the fastest and most accurate method \cite{sivers-2013,sivers-2013b}. For our applications, the method never failed to converge with only five starting simplices, one at the centre of the disc and four rotating on the circle around the pedestrian. Other optimisation methods, for example the Powell algorithm \cite{powell-1964} or gradient based methods might speed up the simulation. However, for our purpose, the Nelder-Mead simplex algorithm and a maximum tolerance of $0.01m$ suffices. Fig.~\ref{fig:timemeasurement} shows the run time for simulations in a long corridor in dependence of the number of people, and thus, the density. We simulated more than 400 people in real time in dense situations. Online visualisation and data recording for later analysis were switched on. We used a standard laptop with Intel\textsuperscript{\textcopyright} Core\textsuperscript{\texttrademark} i7-2670QM CPU, 2.2 GHz, 16 GB RAM and 64 bit Windows 7 operating system. 

\section{Calibration and Validation \label{validation}}  

To validate a model means to show that it matches reality in scenarios relevant to achieve the simulation goal. The model can never be proven right, but only prove its mettle for its application.
In our case we wish to build models that give qualitatively and, to a certain extent, quantitatively realistic predictions of ineffective, in terms of traffic management, or even dangerous situations within a moving crowd. In particular, high crowd densities should be well resolved. When safety of life and limb is at task, we feel that validation must be handled strictly.
Hence, within our VADERE simulation platform, every model we investigate is checked against the tests listed in the RiMEA validation guideline \cite{rimea-2009} with the exception of those test where we do not claim to model the situation, e.g., movement on several levels of a building. 
Not only the OSM is tested in this way but also the implementations of a social force model and a cellular automaton model that we use for comparison in \cite{seitz-2012, dietrich-2013, sivers-2013b, dietrich-2014b}.

Even without an empirically tested model of personal space, navigation becomes more realistic in the sense that pedestrian can make small evasive steps and no longer get stuck in difficult navigational situations. In \cite{sivers-2013b}, pedestrians without the ability to adapt their stride length fail to pass a column placed in the middle of a narrow corridor, reducing flow to zero, while pedestrians with stride adaptation succeed. In this paper we go beyond visual validation and investigate the combined impact of stride adaptation and Hall's personal space model for two special scenarios where experimental data is available:
\begin{itemize}
	\item The density-speed relationship expressed through fundamental diagrams. We use experimental data to calibrate our model parameters. Furthermore, we go beyond the qualitative match required in the RiMEA guideline test 10. We demand that the model is capable of calibrating to any measured density-speed relation.
	\item Behaviour and densities at a bottleneck. We show that pedestrians in a dense crowd make smaller steps and that the higher densities in front of the bottleneck can be reproduced by the improved OSM. Moreover, we demonstrate the impact of the personal space model. We base our scenario on the experiments from \cite{seyfried-2009,seyfried-2009b,liddle-2011,liddle-2011b}.
\end{itemize}

\subsection{The density-speed relation \label{subsec:density-speed}}

The density-speed or density-flow relation, expressed through fundamental diagrams, is a standard benchmark for the simulation of pedestrian traffic: The denser the crowd, the slower it moves. The precise shape of the fundamental diagram depends on the context of the situation. E.g. do we model the rush hour at a German railway station or tourist traffic in, say, India?   
We refer to \cite{seyfried-2005,helbing-2007,moussaid-2009b,schadschneider-2011b,davidich-2012,davidich-2013} to see the present state of the discussion and conclude that we need the possibility to flexibly calibrate the model to any measured fundamental diagram \cite{chattaraj-2009,davidich-2012}. 

The numerical experiment follows the suggestions for test 10 in the RiMEA validation guideline \cite{rimea-2009}. Pedestrians walk through a corridor of length $30m$ and width $4m$. At the end they are transported back to the start so that a perfect loop, without artefacts from corners, is formed. Mathematically this corresponds to periodic boundary conditions. The idea behind this is that, for each initial number of pedestrians in the corridor, a steady state solution should emerge where density and speed no longer change over time. This allows us to measure speed and density without worrying about local effects, at least for the numerical experiment. 

We would like to point out that the same can never be expected for life measurements. That underlines the importance of calibration to any measured fundamental diagram within an acceptable error margin over accurate calibration to a particular diagram. Please refer to \cite{davidich-2013} for a more in-depth discussion on robust calibration and, also, the need for sensitivity studies.  

In state-of-the-art models pedestrians slow down with increasing density. However, the effect is often overestimated and a realistic density of, say  $5 \, \frac{{persons}}{m^2}$ before the crowd comes to a complete halt, may never be reached. If one reduces the torso diameter, thus allowing a tighter packing of the crowd, the effect would be underestimated resulting in a curve above the fundamental diagram. Social force type models \cite{hoecker-2009} and cellular automata \cite{klein-2010,davidich-2012} can use an extra speed adjustment algorithm to fit a given density-speed curve. This can also be done for the OSM with optimisation on a circle \cite{seitz-2012}.
In all cases pedestrians are slowed down according to the density measured during the simulation run when the speed of the crowd surpasses the one given in the fundamental diagram, thus imposing the fundamental diagram from the outside. It is open to debate whether such an imposition is justified. Maybe the most significant advantage of the OSM with optimisation on a disc is that we are capable of calibrating it by adjusting the model parameters that deal with the need for personal space alone. 

\begin{figure}
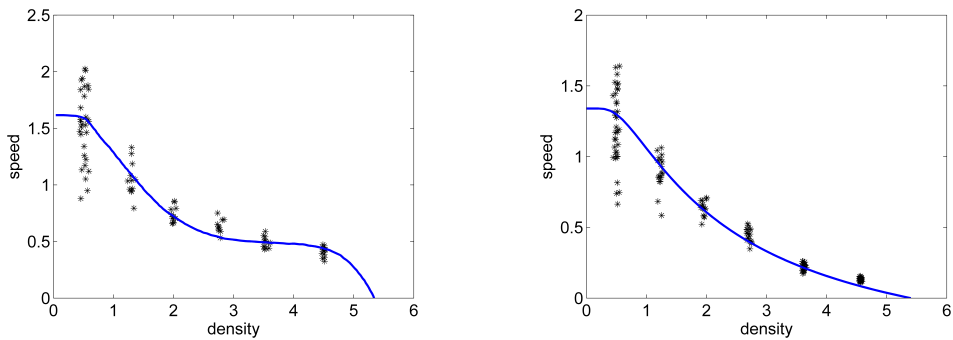

\centering
	\subfigure[Simulation output (black asterisks) after calibrating the social parameters to the reference curve (blue solid line) of Hankin and Wright \cite{hankin-1958}. The resulting parameters are $\mu_p = 30.0$, $a_p = 2.0$ and $b_p =4$.]{
	\label{fig:Hankin}\includegraphics[width=0.45\textwidth]{Hankin.pdf}}
	\hspace{0.5cm}
	\subfigure[Simulation output (black asterisks) after calibrating the social parameters to the reference curve (blue solid line) of Weidmann \cite{weidmann-1993}. The resulting parameters are $\mu_p = 50.0$, $a_p = 1.2$ and $b_p = 1$.]{
	\label{fig:Weidmann}\includegraphics[width=0.45\textwidth]{Weidmann.pdf}}
	\caption{Results of simulations with different densities in the OSM. Calibration to different fundamental diagrams is possible.}
	\label{fig:fundDIA}
\end{figure} 
 
With the OSM with optimisation on a disc, pedestrians react to a dense situation by making smaller, hesitant or adaptive steps (empirically observed, e.g., in \cite{jelic-2012b}) that lead to a speed reduction \cite{seyfried-2010}. 
Adjustment of the parameters $\mu_p$, $a_p$ and $b_p$ for `pedestrian 
avoidance' suffices to achieve a good quantitative fit to a given fundamental diagram.
In Fig.~\ref{fig:Hankin} and Fif.~\ref{fig:Weidmann}, we show results for two different flow types: the fundamental diagram by Hankin \cite{hankin-1958} for pedestrian flow in a subway and the benchmark diagram by Weidmann suggested in the RiMEA test guideline \cite{weidmann-1993,rimea-2009}. See Fig.~\ref{fig:Weidmann}) for Hankin and Wright's diagram and Fig.~\ref{fig:Hankin}) for Weidmann's diagram. The calibration can be done by simulating several runs (see \cite{seitz-unpublished}) with different parameters or, more elegantly, by solving a mathematical minimization problem (see \cite{davidich-2012}). For Hankin and Wright we obtain the following social parameters: $\mu_p = 30.0$, $a_p = 2.0$ and $b_p = 4$ with a mean free-flow speed of 1.62 $\frac{m}{s}$ and variance 0.26 $\frac{m}{s}$; For Weidmann we get: $\mu_p = 50.0$, $a_p = 1.2$ and $b_p = 1$ with a mean free-flow speed of 1.34 $\frac{m}{s}$ and variance 0.26 $\frac{m}{s}$.

This matches well with the idea that the differences between fundamental diagrams are caused by physical variations, such as the average fitness of people forming a crowd, as well as social and cultural variations in the need for personal space. There may be many more sources of influence. 
However, we believe the main variations to be captured in the fundamental diagrams. Density-speed relationships measured in different countries show different curves (see \cite{older-1968,weidmann-1993,lam-1995}). Indian students, for example, showed a less pronounced need for personal space and strength of avoidance than German students in identical experiments described in \cite{seyfried-2005,chattaraj-2009}. Also the fundamental diagrams measured at a major German railway station differed for a slow and, perhaps, tired early morning crowd and for faster commuters heading home in the evening rush hour \cite{davidich-2013}.

\subsection{Behaviour at a bottleneck \label{subsec:bottleneck}}

We look at a bottleneck scenario where experimental data is available \cite{seyfried-2010,seyfried-2010b,liddle-2011,liddle-2011b}. Our goal is to demonstrate the impact of the stride length adaptation and of the new personal space model in the OSM. In particular, we show that the variation of the densities that is observed in front of the bottleneck can be qualitatively, and to a certain extent, even quantitatively reproduced with the improved OSM. Here, we focus on comparisons with the former state of the OSM, rather than on comparisons to cellular automata that cannot produce fine variations in density or on comparisons to continuous models that do not model steps. 
Unfortunately, the experimental data on bottlenecks does not come with the corresponding free-flow speed measurements and fundamental diagram for all densities. Hence, we produce the first results with a free-flow speed of 1.62 $\frac{m}{s}$ and parameters $\mu_p = 30.0$, $a_p = 6.0$ and $b_p 70= 4$. 
The first two values would be typical for fit young subjects which one would hope to be the case for the soldiers in \cite{liddle-2011}.
Thus, we are able to obtain qualitative meaningful results but cannot expect a close quantitative match. To exactly reproduce the behaviour at the bottleneck, we would need more data to calibrate our model.

\begin{figure}
\centering
	\subfigure[OSM without stride adaptation: All steps have equal length.]{
	\label{fig:bottleneck_D100_traj}\includegraphics[width=0.4\textwidth]{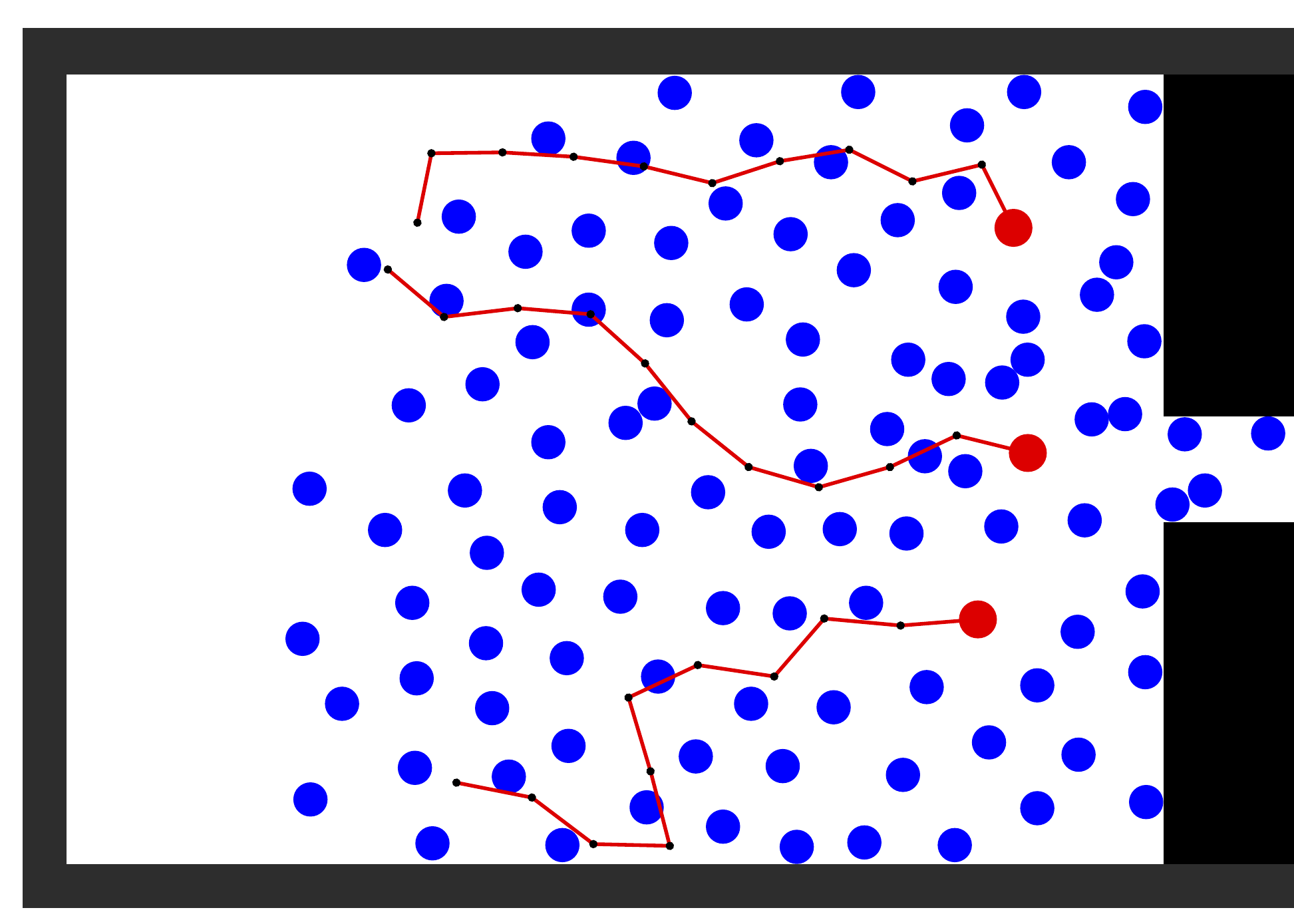}}
	\hspace{0.5cm}
	\subfigure[OSM with stride adaptation: Each pedestrian adjusts the stride length according to the situation.]{
	\label{fig:bottleneck_NM_traj}\includegraphics[width=0.4\textwidth]{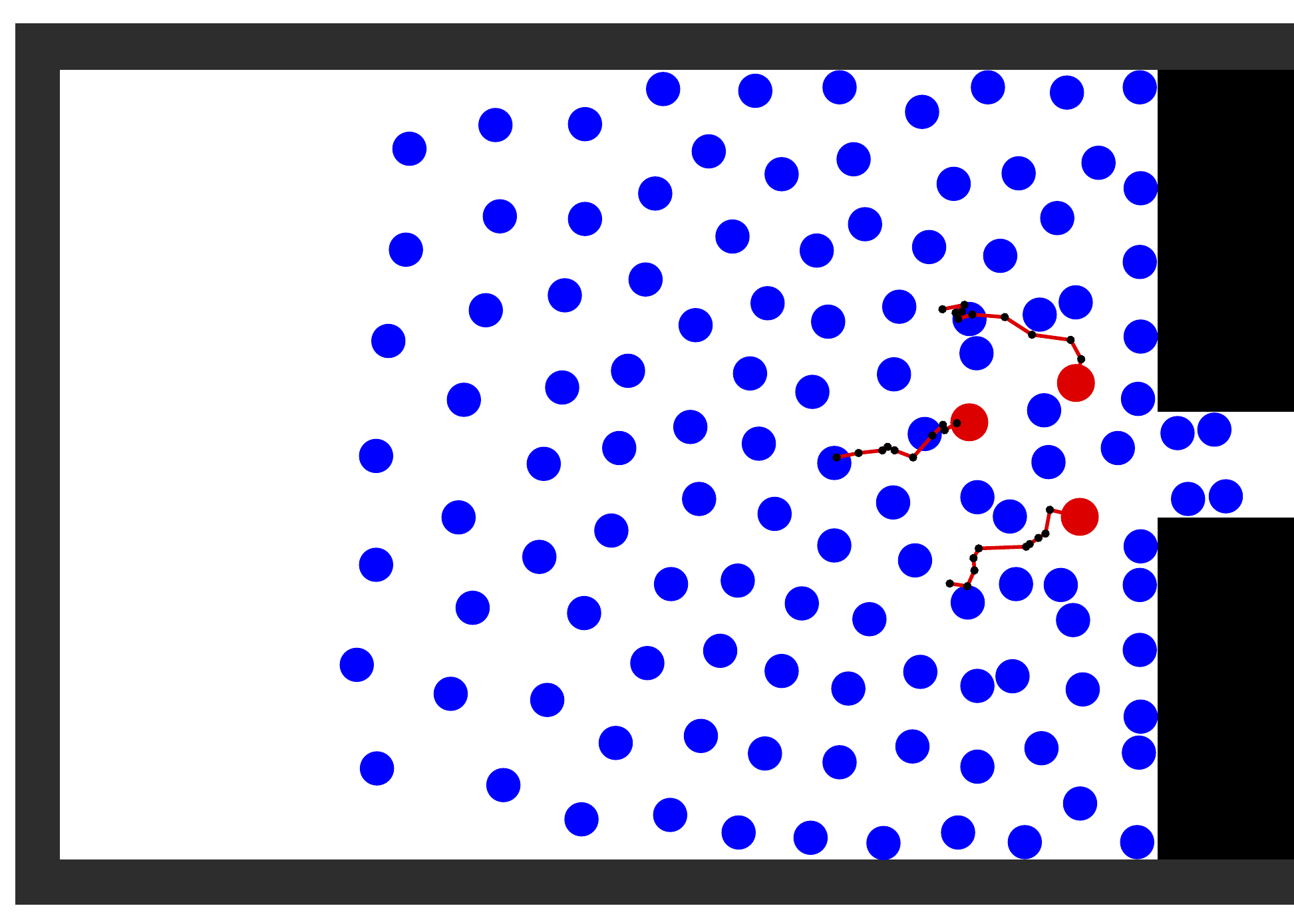}}
	\subfigure[OSM without stride adaptation: No clear regions of increasing densities can be observed in front of the bottleneck.]{
	\label{fig:bottleneck_D100_dens}\includegraphics[width=0.4\textwidth]{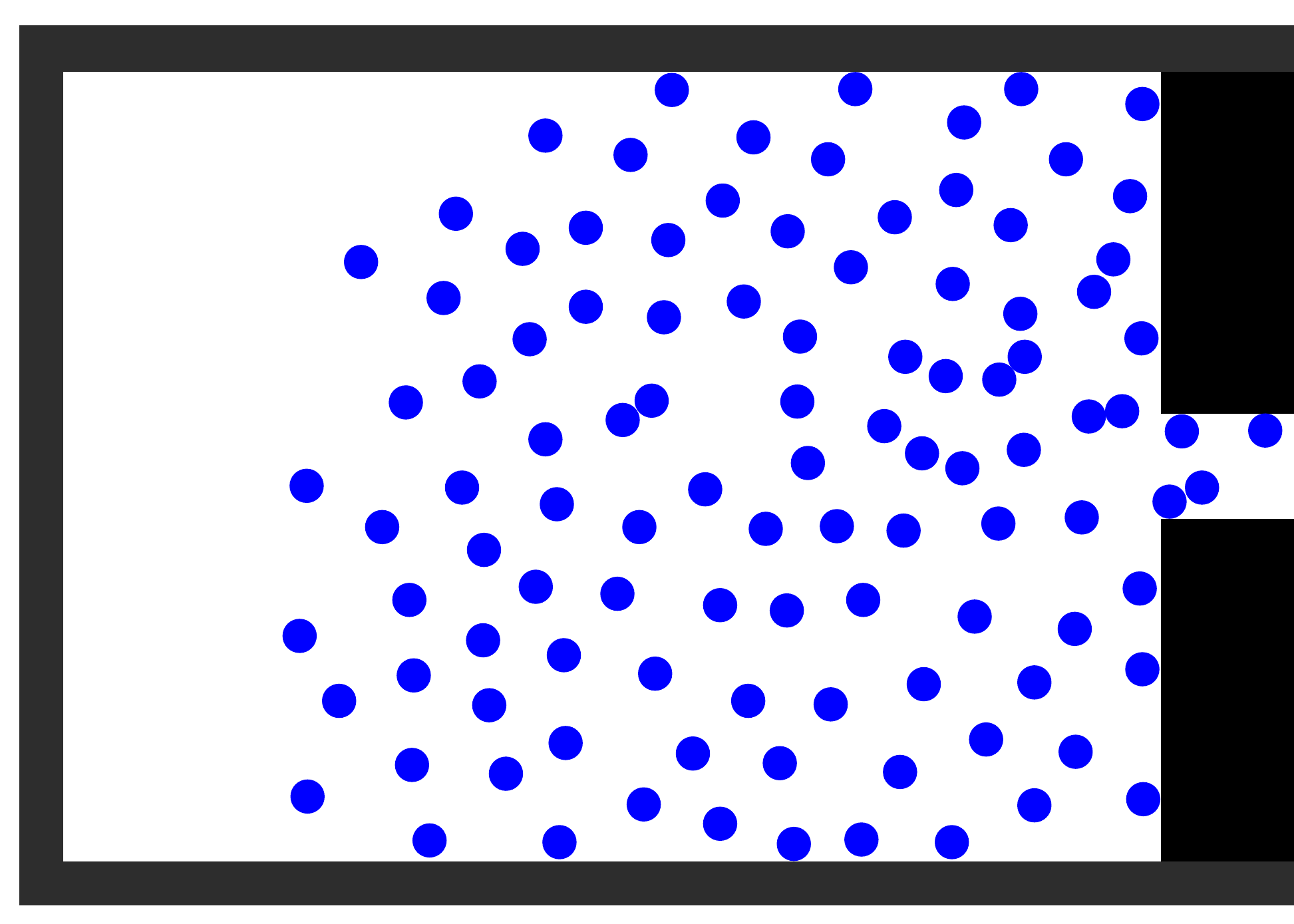}}
	\hspace{0.5cm}
	\subfigure[OSM with stride adaptation: In front of the bottleneck, the density is higher than 2.0 $\frac{pers}{m^2}$ and decreases towards the back.]{
	\label{fig:bottleneck_NM_dens}\includegraphics[width=0.4\textwidth]{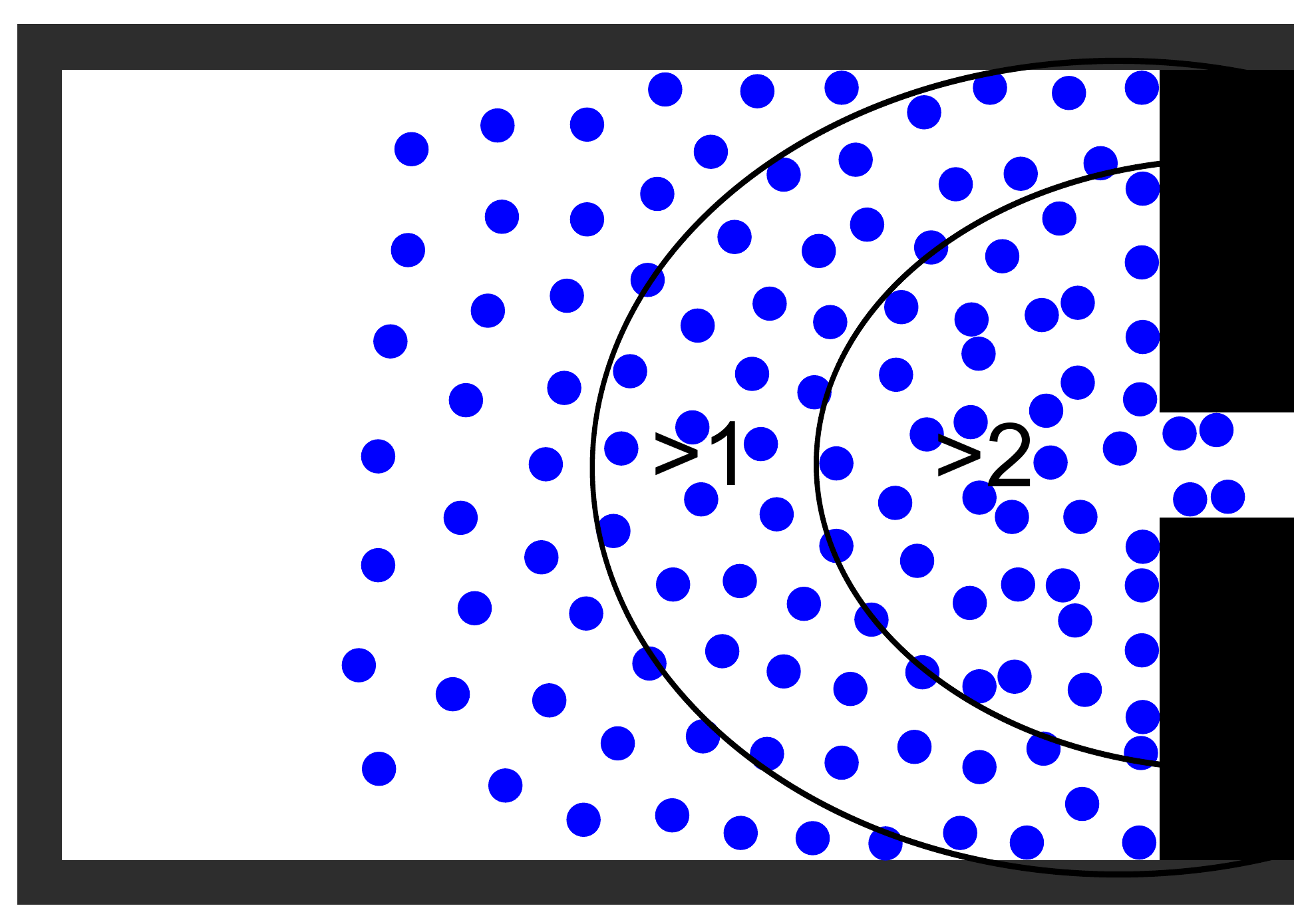}}
	\subfigure[OSM without stride adaptation: Preservation of the personal spaces even fails at low densities.]{
	\label{fig:bottleneck_D100}\includegraphics[width=0.4\textwidth]{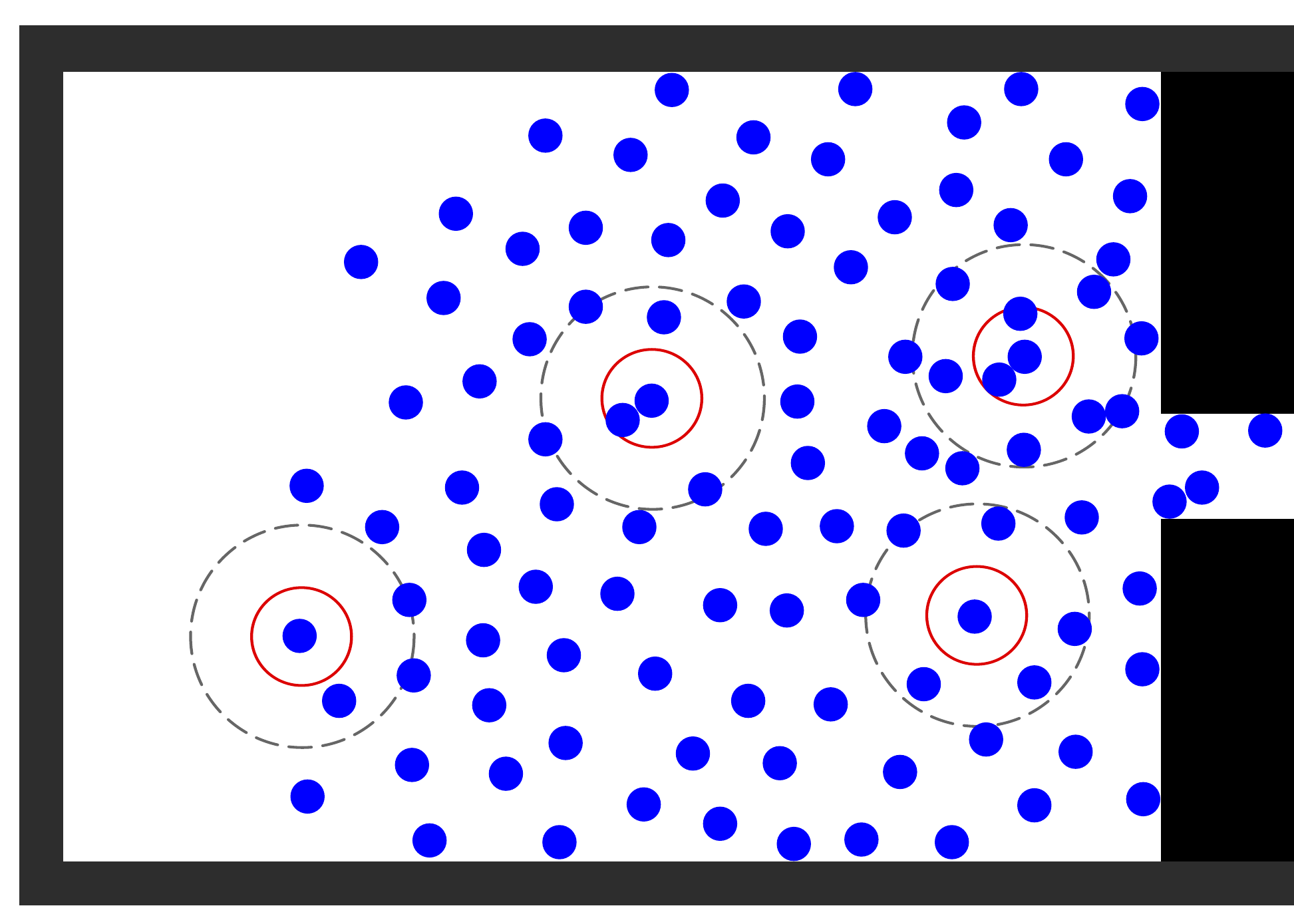}}
	\hspace{0.5cm}
	\subfigure[OSM with stride adaptation: Pedestrians keep their intimate space clear until it is too dense.]{
	\label{fig:bottleneck_NM}\includegraphics[width=0.4\textwidth]{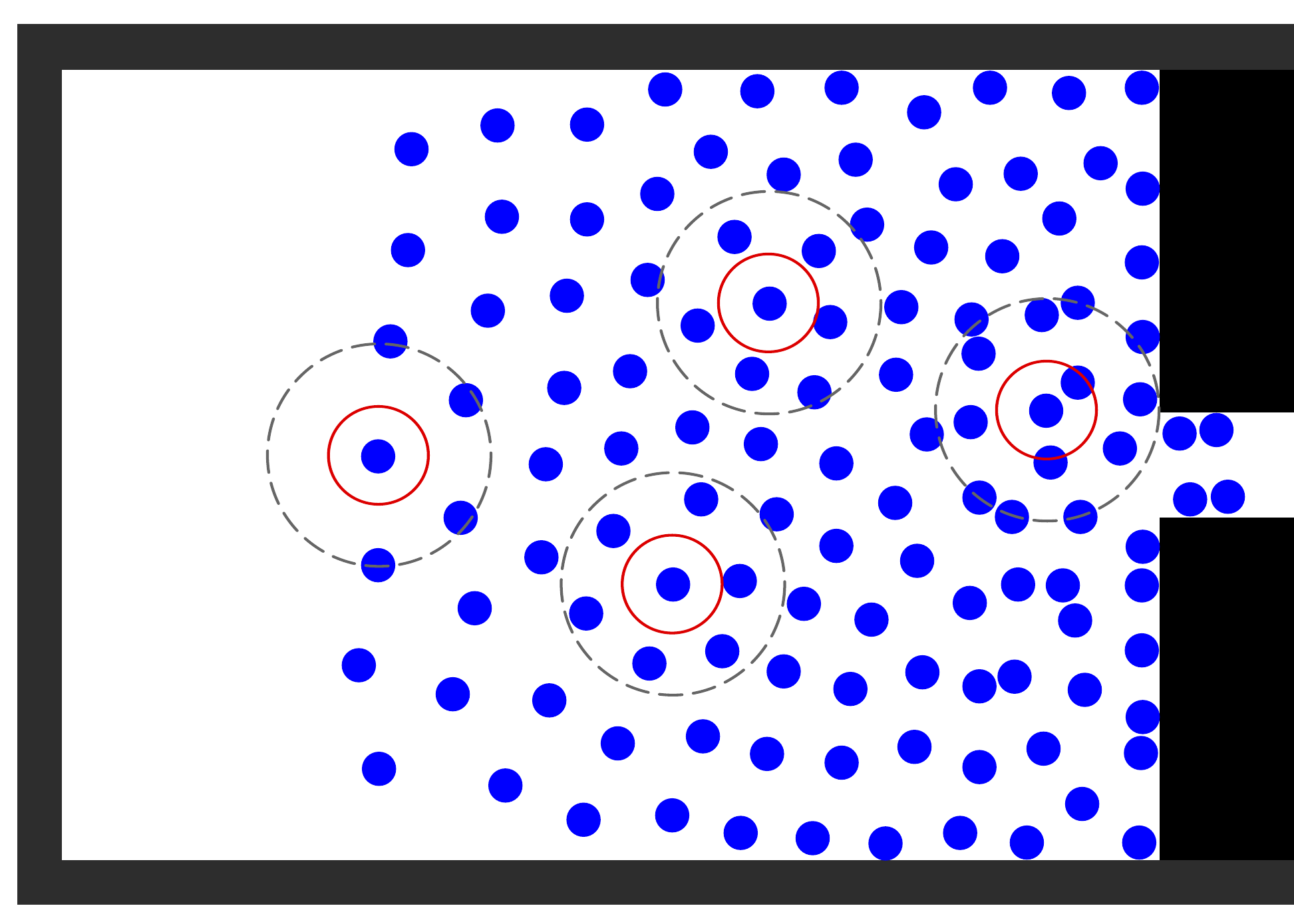}}
	\caption{Screenshots during a bottleneck simulation with the OSM. In a) and b), the last ten strides of three arbitrary pedestrians are marked with (red) lines and separated with (black) dots. In c) and d), we look at the densities in front of the bottleneck. Finally, in e) and f), we mark the modelled spaces around some pedestrians.
	For the calibration parameters refer to tables \ref{tbl:parametersEXP} and \ref{tbl:parametersCAL}.} 	\label{fig:bottleneck}
\end{figure}

In dense situations, e.g., in front of a bottleneck \cite{liddle-2011,liddle-2011b}, people tend to make smaller steps. With the optimisation on the whole disc in the OSM, we can observe this behaviour in our outcomes. In Fig.~\ref{fig:bottleneck_NM_traj} and \ref{fig:bottleneck_D100_traj}, the ten last steps of three arbitrary pedestrians in front of the bottleneck are marked. With the optimisation on the circle, the pedestrians only take large steps. They often skip a step, thus slowing down. 
Without stride adaptation, pedestrians need more time to take ten steps in dense situations than in free-flow situations. It is not possible for the pedestrians to gradually move closer to others. With the optimisation on the disc, pedestrians make smaller steps in dense situations. They move up behind the pedestrians in front of them, make evasive steps to increase their free personal space and adaptive steps to navigate through the other pedestrians. 
The pedestrians are distributed more evenly in space.

Smaller steps also enable denser crowds. Without stride adaptation the OSM produces densities in front of the bottleneck that do not increase significantly. See Fig.~\ref{fig:bottleneck_D100_dens}. With the optimisation on the disc, that is, with stride adaptation, we observe higher densities in front of the bottleneck as reported in \cite{liddle-2011,liddle-2011b}. See Fig.~\ref{fig:bottleneck_NM_dens}.

Following the personal space model from Hall \cite{hall-1966}, humans try to keep their intimate space clear from strangers. Only in dense situations, when people cannot avoid having others in their nearest neighbourhood, or when they feel a shared social identity (e.g. \cite{novelli-2013,alnabulsi-2014}), do they accept strangers in their intimate space. This behaviour should also be observed in simulations of pedestrian movement. 
With the new personal space model in the OSM, the effect can be clearly seen (see Fig.~\ref{fig:bottleneck_NM}). In situations with a low density, smaller than 1.0 $\frac{pers}{m^2}$, most pedestrians can keep their intimate space clear of others. They even preserve the close zone of their personal space. Between 1.0 $\frac{pers}{m^2}$ and 2.0 $\frac{pers}{m^2}$ pedestrians still keep their intimate space clear. Strangers only enter this zone if they overtake. In denser situations, starting at about 2.0 $\frac{pers}{m^2}$, the intimate space cannot be kept clear any more. Without stride adaptation in the OSM (see Fig.~\ref{fig:bottleneck_D100}), pedestrians must accept others in their personal and intimate spaces at much lower densities. We conclude that the the personal space model needs a fine resolution in the locomotion model to unfold its potential.

\begin{figure}
\centering
	\subfigure[Screenshot in front of the bottleneck after 16 seconds. The (red) rectangle marks the density measurement area (1m$\times$1m).]{
	\label{fig:bottleneck_density}\includegraphics[height = 5.2cm]{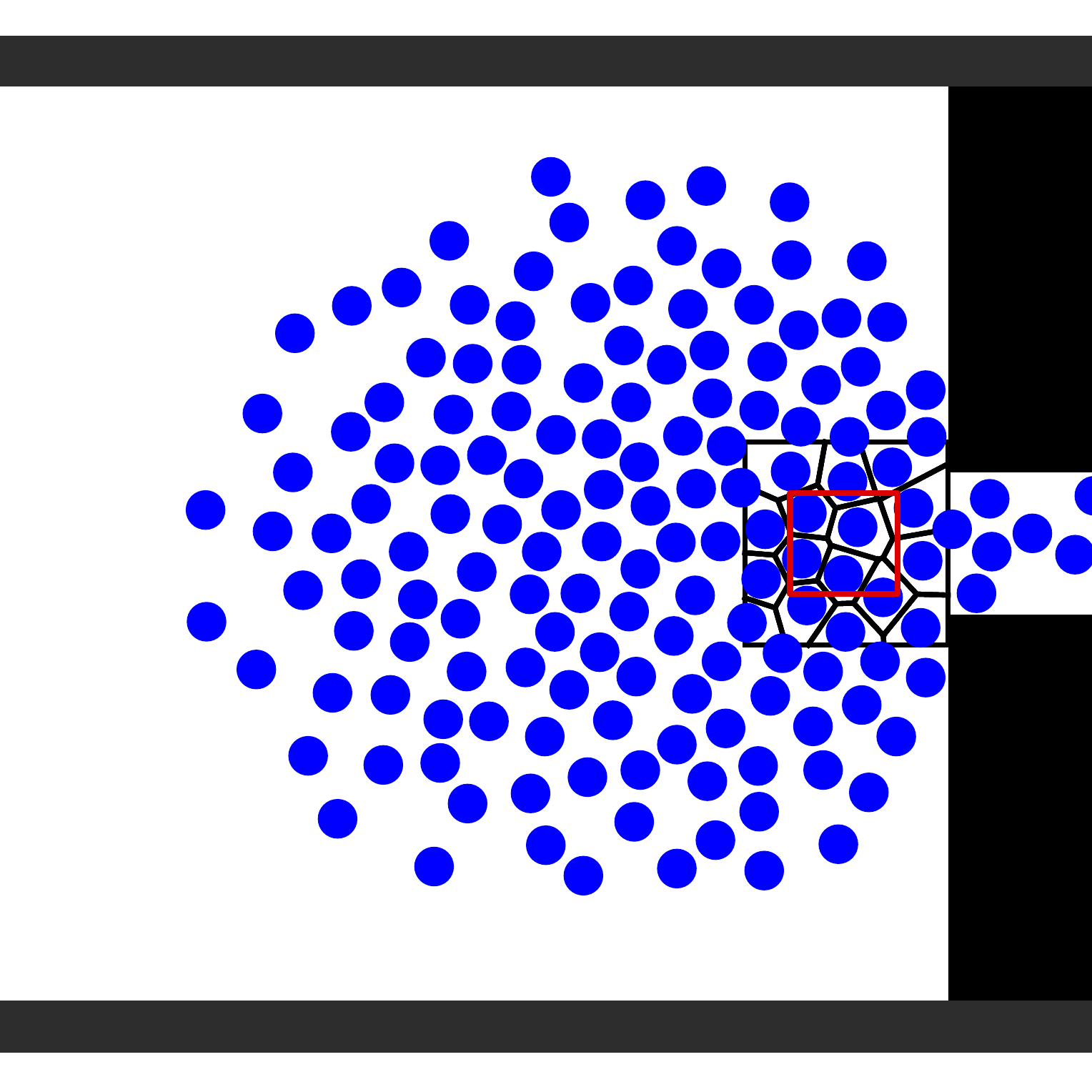}}
	\hspace{0.5cm}
	\subfigure[Measured densities in front of the bottleneck. Density is measured using Voronoi cells during the simulation. The result is close to the corresponding one from the experiment in \cite{liddle-2011}.]{
	\label{fig:bottleneck_simtime}\includegraphics[height = 5.2cm]{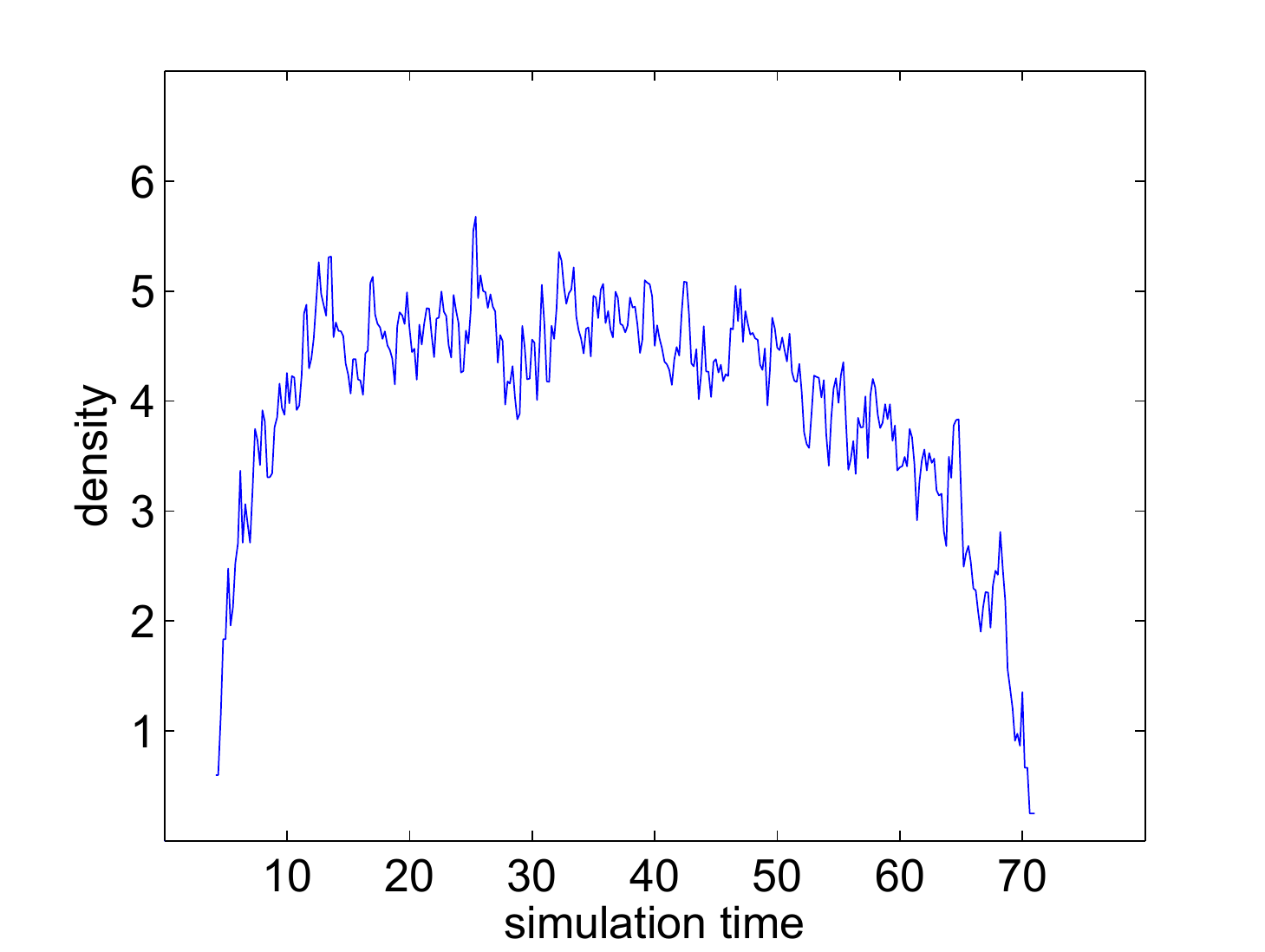}}
	\caption{Simulation of a bottleneck scenario (width = 1.4m, length = 4m) from \cite{liddle-2011}. Pedestrians move from left to right and form a loose queue induced by the density dependent floor field from \cite{zoennchen-2013,koster-2014b}. The parameters for personal space are calibrated to the experiments in \cite{liddle-2011,liddle-2011b}
	and listed in tables \ref{tbl:parametersEXP} and \ref{tbl:parametersCAL}. In accordance with the experiment, densities up to 5 $\frac{pers}{m^2}$ occur in front of the bottleneck.}
	\label{fig:bottleneck_densities}
\end{figure}

However, the densities in front of the bottleneck in Fig.~\ref{fig:bottleneck_NM_dens} are still lower than the measurements reported in \cite{liddle-2011}. Also the queue forms generated by the 180 participants in \cite{liddle-2011} suggest that the subjects loosely queued up when leaving the room instead of jostling for the closest position to the opening. In our model, this type of behaviour can be induced by a dynamic floor field that makes densely populated areas attractive. 
See \cite{zoennchen-2013,koster-2014b} for a detailed discussion.

To exactly reproduce the data from \cite{liddle-2011}, we would need the size of the bodies, the free-flow speed of the 180 participants or the fundamental diagram. We do not have this but still want to demonstrate the
ability to match the compression in front of a bottleneck.
Thus we must make reasonable assumptions. We believe that the soldiers in the experiments of \cite{seyfried-2010,seyfried-2010b,liddle-2011,liddle-2011b} have a larger body than average persons. Thus, we assume a radius of 0.22m for our simulation.
Furthermore, we set the free-flow speed in our simulation to 1.2~$\frac{m}{s}$. This is the average speed measured
for soldiers in a very low density situation (0.25~$\frac{persons}{m^2}$) in the same series of experiments as \cite{liddle-2011} and reported in \cite{seyfried-2010}.
We use a dynamic floor field where the dense crowd has an attractive effect. A queue forms. The width is 
controlled by parameter $c_q=0.5$ that goes into the right hand side of the eikonal equation (\ref{eqn:eikonal}). See \cite{koster-2014b, zoennchen-2013} for details. This time, we calibrate to the measured evacuation times from \cite{liddle-2011} instead of a fundamental diagram and get $\mu_p = 5.0$, $a_p = 1.0$ and $b_p = 1$. 
The parameters correspond to a low need for personal space which we find plausible for soldiers trained to march together.
With these parameters we correctly reproduce the highest density area in front of the bottleneck of width 1.4m, length 4.0m in \cite{liddle-2011}. That is, the compression is in the right location and the density value is about 5~$\frac{persons}{m^2}$ as in the experiment.
We also reproduce the steady state phase of the densities. See Fig.~\ref{fig:bottleneck_density} and  Fig.~\ref{fig:bottleneck_simtime}.

\begin{table}
	 \centering
		\begin{tabularx}{0.95\textwidth}{|p{0.15\textwidth}|X|c|l|}
			\hline
			Parameter & Description & Value & Experiment\\
			\hline
			$\delta_{int}$ & intimate distance & 0.45 m & \cite{hall-1966}\\
			$\delta_{pers}$ & personal distance & 1.20 m & \cite{hall-1966}\\
			$\delta_o$ & distance kept from obstacles & 0.8 m & \cite{seitz-unpublished}\\
			\hline
		\end{tabularx}
		  \caption{Parameters for the `pedestrian avoidance' and `obstacle avoidance' in the OSM that are taken from experiments.}
     \label{tbl:parametersEXP}
\end{table}

\begin{table}
	 \centering
		\begin{tabularx}{0.95\textwidth}{|p{0.15\textwidth}|X|c|c|c|c|}
			\hline
			& & \multicolumn{4}{c|}{Values}\\
			Parameter & Description & 1 & 2 & 3 & 4  \\
			\hline
			$\mu_p$ & strength of `pedestrian avoidance' & 30.0 & 50.0 & 30.0 & 5.0 \\
			$a_p$ & moderation between intimate and personal space & 2.0 & 1.2 & 6.0 & 1.0 \\
			$b_p$ & intensity of transition between intimate and personal space & 4 & 1 & 4 & 1 \\
			$\mu_o$ & strength of ´obstacle avoidance' & 6.0 & 6.0 & 6.0 & 6.0 \\
			$c_q$   & governs the queue width          & --  & --  & --  & 0.5 \\
			\hline
		\end{tabularx}
		  \caption{Calibrated parameters for the `pedestrian avoidance' and `obstacle avoidance' in the OSM. Scenarios: 1 = fundamental diagram Hankin\&Wright \cite{hankin-1958}, 2: fundamental diagram Weidmann \cite{weidmann-1993}, 3:  Bottleneck - comparison former and new OSM, 4: Bottleneck - reproducing Liddle et al.\cite{liddle-2011}.}
     \label{tbl:parametersCAL}
\end{table}

\section{Discussion and outlook \label{conclusion}}

Pedestrian motion operates on a strategic, a tactical and an operational level \cite{hoogendoorn-2004}: When and where to do pedestrians decide to walk? How do they find their paths? How do they step forward? 
The Optimal Steps Model is a model for locomotion, that is, it describes the operational level.
We believe that an inaccurate model of the operational level corrupts results on the higher levels. Examples are inaccuracies and oscillations as seen  in some continuous models 
(see e.g. \cite{chraibi-2010} or \cite{koster-2013}) that may cause people to get stuck, lose their way, or hurry back and forth in a way that has no relation to meaningful strategic or tactical decisions. Grid restrictions as in cellular automata \cite{koster-2011, seitz-2012} are just as undesirable because they limit the choice of directions and make it impossible or, at least, extremely difficult to map a gradual compression of a crowd \cite{was-2013}.
The problem goes even deeper. Humans neither move on smooth rails, nor do they hop from cell to cell. Instead they step forward in continuous space while instantly adapting their stride length and speed to the navigational situation. 

In this paper we presented an extension of the  Optimal Steps Model that exactly maps this natural behaviour. We believe the consequences to be considerable:
With stride adaptation it made sense, for the first time, to incorporate an empirically tested, and finely resolved, psychological model of personal space \cite{hall-1966}. 
As a result calibration to different measured density-speed or, equivalently, density-flow relationships was achieved simply by adjusting the need for personal space. This implies that flow was modelled correctly.
Also, the simulation results for important benchmark scenarios, such as navigation around a column in a narrow corridor \cite{sivers-2013b} and a crowd in front of a bottleneck, matched reality much more closely than the former state-of-the-art. We believe that this increases the predictive power of pedestrian traffic models.

The mathematical formulation came in the form of a two-dimensional optimisation problem that was solved successfully, and efficiently, with standard numerical algorithms. From a numerical analyst's point of view the accuracy and speed of the simulation might be further improved, e.g., by using optimisation algorithms that need first derivatives of the objective function.

However, our main goal was to provide a sound operational model that is based on empirically substantiated psychological findings and matches human locomotion to a significantly higher degree that the former state-of-the-art.
Our vision is to incorporate sociological and further psychological aspects into our model.
 
\section*{Acknowledgments}

This work was funded by the German Federal Ministry of Education and Research through the projects MEPKA on mathematical characteristics of pedestrian stream models (Grant No. 17PNT028) and MultikOSi on assistance systems for urban events(Grant No. 13N12824).





\bibliographystyle{elsarticle-harv}
\bibliography{Literature}







\end{document}